\documentclass[a4paper]{jpconf}
\usepackage{graphicx}
\usepackage{amsmath}
\usepackage{amssymb}
\usepackage{color}
\begin{document}
\title{Finite-size scaling of Lennard-Jones droplet formation at fixed density}

\author[]{Johannes Zierenberg and Wolfhard Janke}
\address{Institut f\"ur Theoretische Physik, Universit\"at Leipzig, Postfach 100\,920, 04009 Leipzig, Germany}

\ead{johannes.zierenberg@itp.uni-leipzig.de, wolfhard.janke@itp.uni-leipzig.de}

\begin{abstract}
  We reaccess the droplet condensation-evaporation transition of a
  three-dimensional Lennard-Jones system upon a temperature change. With the
  help of parallel multicanonical simulations we obtain precise estimates of
  the transition temperature and the width of the transition for systems with
  up to $2048$ particles. This allows us to supplement previous observations of
  finite-size scaling regimes with a clearer picture also for the case of a
  continuous particle model.
\end{abstract}

\newcommand{\kB}{k_{\rm B}}

\vspace*{-8pt}
\section{Introduction}
Despite the long-lasting scientific interest in droplet formation, it is still
a modern problem with many open questions. This is partially due to the general
nature of droplet formation, with relevance ranging from metastable decay over
cloud formation to cluster formation in protein solutions. In this work, we
consider equilibrium droplet formation which yields a firm basis to study the
transition between a homogeneous gas and a mixed phase of a droplet in
equilibrium with surrounding vapor~\cite{binder1980, neuhaus2003, biskup2002,
biskup2003, binder2003}. The problem is formulated in the canonical ensemble. A
common control parameter is the density while the temperature is fixed. This
allows one, in principle, to access temperature-dependent quantities such as
the isothermal compressibility and the interface tension.

At fixed temperature the theory has been verified by several computational
studies, including lattice~\cite{nussbaumer2006, nussbaumer2008,
zierenberg2014jpcs} and Lennard-Jones~\cite{macdowell2004, macdowell2006,
schrader2009} systems. Instead, however, one can fix the density and vary the
temperature which yields an equivalent but ``orthogonal'' finite-size scaling
behavior~\cite{martinos2007,zierenberg2015}. In this case, we recently observed
an intermediate scaling regime consistent with finite-size scaling results for
polymer aggregation~\cite{zierenberg2014jcp}. Here, we present new results for
the three-dimensional Lennard-Jones system at fixed density which extend and
supplement our previous results~\cite{zierenberg2015}. 

\section{Model and Method}
\begin{figure}
  \vspace{-1em}
  \centering
  \includegraphics[width=0.45\textwidth]{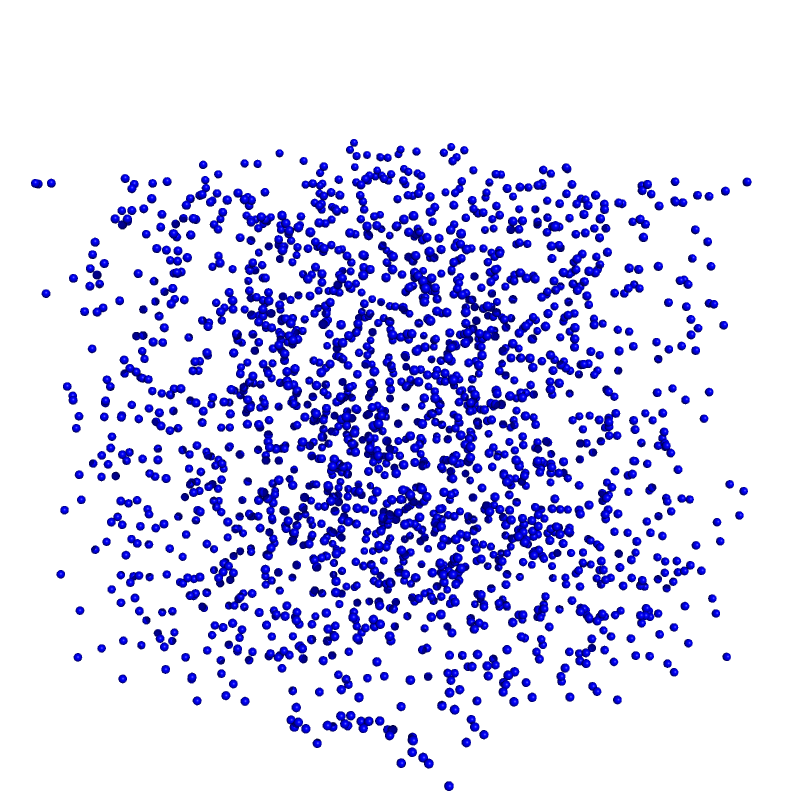}
  \hspace{1em}
  \includegraphics[width=0.45\textwidth]{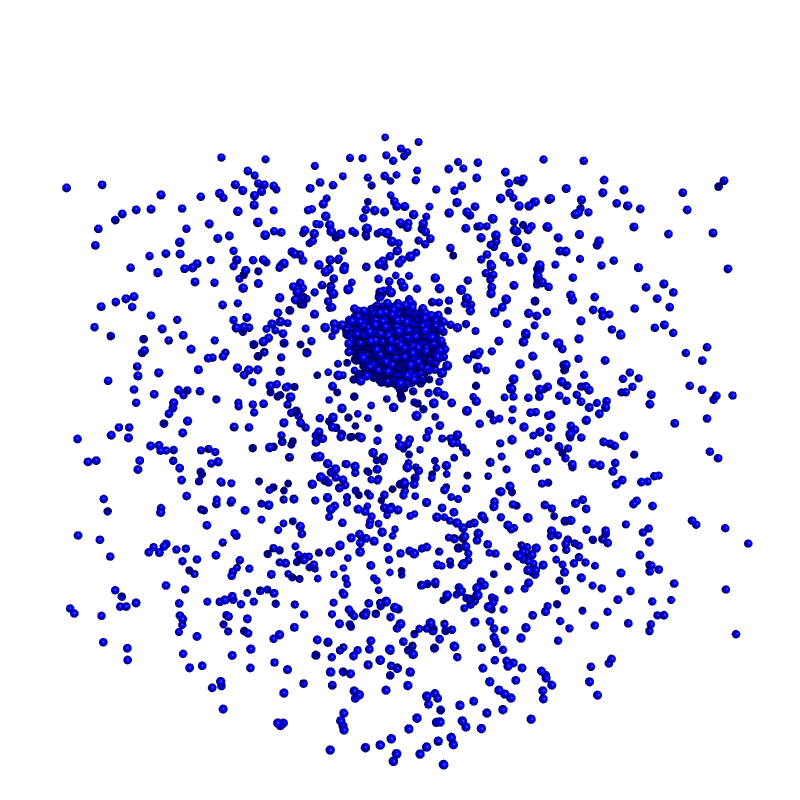}
  \caption{\label{figSetup}%
    Illustration of a three-dimensional Lennard-Jones gas (left) and a droplet
    in equilibrium with surrounding vapor (right). Shown is a system with $N=2048$
    particles at density $\rho=10^{-2}$ in a box with periodic boundary
    conditions. 
  }
\end{figure}

We consider $N$ Lennard-Jones particles in a three-dimensional box of length $L$
with periodic boundary conditions, see Fig.~\ref{figSetup}. The self-avoidance
and short-range attraction is modeled by the pairwise Lennard-Jones potential 
\begin{equation}
  V_{\rm LJ}(r_{ij}) 
      = 4\epsilon\left[\left(\frac{\sigma}{r_{ij}}\right)^{12} 
      - \left(\frac{\sigma}{r_{ij}}\right)^6\right],
\end{equation}
where $r_{ij}$ is the distance between particle $i$ and $j$. As in
Ref.~\cite{zierenberg2015}, we set $\epsilon=1$ and $\sigma=2^{-1/6}$. The
computational demand can be reduced by introducing a cutoff radius
$r_c=2.5\sigma$ above which particles do not interact anymore. The potential is
then shifted by $V_{\rm LJ}(r_c)$ in order to be continuous, yielding
\begin{equation}
  V^*_{\rm LJ}(r) = 
  \begin{cases}
    V_{\rm LJ}(r)-V_{\rm LJ}(r_c) & r < r_c \\
    0                             & \mathrm{else}
  \end{cases}.
\end{equation}
This is in accordance with the existing literature and enables the application
of a domain decomposition, where the periodic box is decomposed into equally
large (cubic) domains. These domains have to be at least of the size $r_c$.
Then, the interaction of each particle is obtained by evaluating only its domain
and the adjacent ones (in three dimensions this adds up to $3^3=27$ domains).
Especially in the gas phase the simulation benefits from this procedure, where
the particles are equally distributed in the full box. 

Following Ref.~\cite{zierenberg2015}, we fix the density $\rho=N/L^3$ and vary
the temperature $T$. This allows the application of multicanonical
simulations~\cite{berg1991, berg1992, janke1992, janke1998}, which are
well-suited for first-order phase transitions such as condensation. At a
first-order transition two phases are in coexistence with suppressed transition
states in between. This is circumvented by replacing the canonical Boltzmann weight
$\exp(-E/\kB T)$ with an alterable weight function, which is
iteratively adapted in order to yield a flat histogram in the energy. Each
iteration is in equilibrium, sampling the distribution according to the current
weight function. This leads to a straightforward
parallelization~\cite{zierenberg2013cpc, zierenberg2014pp} which has been shown
to perform very well for the problem at hand~\cite{zierenberg2014jpcs}. In the
end, canonical expectation values are estimated by reweighting the data from a
multicanonical production run.

A crucial aspect is the selection of Monte Carlo updates. While in
Ref.~\cite{zierenberg2015} we restricted ourselves to local particles shifts, we
added here particle ``jumps'' with a larger update range. The update range for a
particle shift was set to $r=0.3$ and for a particle jump to $r=L/2$. A new
position is then selected with equal probability from a sphere of size $r$
around the old position. This simple non-local update significantly enhanced the
sampling of the gaseous phase and increased the performance of the simulation
drastically. Instead of $512$ we are now able to sample up to $2048$ particles.

\section{Theory}
Here we only want to briefly recapture the results of many previous
works~\cite{neuhaus2003, biskup2002, biskup2003, binder2003, zierenberg2015}.
For a supersaturated gas, it was shown that the probability of
intermediate-sized droplets vanishes~\cite{biskup2002, biskup2003}, which leaves
us with the scenario of a homogeneous gas phase on the one side and a droplet in
equilibrium with surrounding vapor on the other side, see Fig.~\ref{figSetup}.
At a fixed temperature, the droplet formation may be achieved by adding more
particle excess to the already supersaturated gas. For both sides it is possible
to formulate a contribution to the free-energy in terms of fluctuations (entropy
of the gaseous phase) and surface tension (energy of the droplet). Importantly,
the finite-size dependence could be rewritten in terms of the fraction of
particle excess in the droplet as a function of a dimensionless density. This in
turn allowed us to expand the results around the infinite-size transition
temperature $T_0$ to yield the leading finite-size scaling behavior of the
condensation temperature $T_c$ at fixed density~\cite{zierenberg2015}.
Similarly, we showed that an expansion of the free-energy difference at the
condensation transition yields an estimate of the leading-order finite-size
scaling of the transition rounding $\Delta T$, i.e., the width of the
transition. For details we refer to the prior literature and give here only the
three-dimensional results to leading order: 
\begin{align}
  T_c -T_0 &\propto N^{-1/4}, \label{eqFSSTC}\\
  \Delta T &\propto N^{-3/4}. \label{eqFSSDT}
\end{align}

A crucial observation is now that the size $R$ of the droplet at the
condensation transition itself scales non-trivially with system size, namely in
three dimensions $R\propto N^{1/4}$. Thus, the leading-order scaling may be
identified in terms of powers of the linear extension of the droplet itself,
$\propto R^{-1}$ respectively $\propto R^{-3}$.
This is consistent with the interpretation that the droplet size at transition
is the relevant system size. Then, a virtual subsystem around the droplet would
lead to a transition between a homogeneous liquid (droplet) phase to a
homogeneous gas phase with open boundary conditions. The competition between
finite-size contributions from volume ($\propto R^3$) and surface ($\propto
R^2$) would in this case give rise to an intuitive finite-size correction of
the order $R^{-1}$~\cite{privman1990, borgs1995, borgs2002, zierenberg2015,
zierenberg2014jcp}.

\section{Results}
\begin{figure}
  \vspace{-1em}
  \centering
  \includegraphics{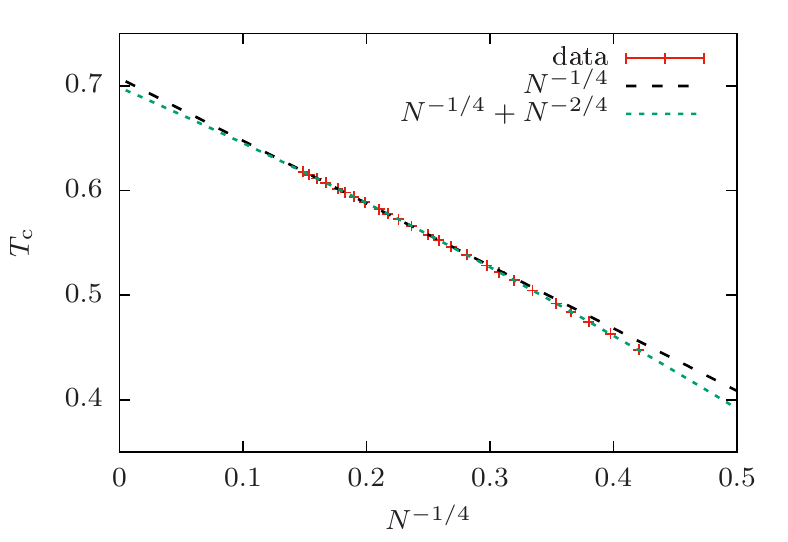}
  \hspace{-1em}
  \includegraphics{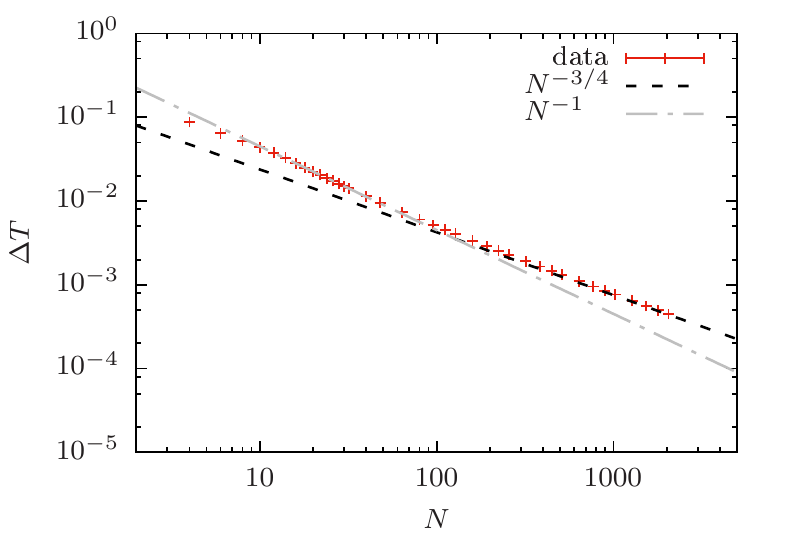}
  \caption{\label{figFSS}%
    Finite-size scaling of the transition temperature (left) and rounding
    (right) as the maximum location and the half width of the specific-heat peak,
    respectively. Error bars are included but smaller than the symbol size. 
  }
\end{figure}

Due to the non-local Monte Carlo update, we are now able to extend our previous
results~\cite{zierenberg2015} for the three-dimensional Lennard-Jones
system to $N=2048$ particles with improved statistics.
The temperature scale is fixed by setting $\kB=1$. The
finite-size transition temperature $T_c$ is obtained as the location of the
largest peak in the specific heat $C_V=\kB\beta^2\left(\langle
E^2\rangle -\langle E\rangle^2\right)/N$ and plotted in Fig.~\ref{figFSS}~(left)
versus the expected scaling behavior Eq.~\eqref{eqFSSTC}. Overall, the data
qualitatively shows a linear behavior as predicted. A leading-order fit for
$N\geq1024$ yields $T_0=0.7032(3)$ with goodness-of-fit parameter $Q\approx0.1$,
shown in the figure as dashed black line. Including additional higher-order
corrections, i.e., $T_c=T_0 +a N^{-1/4}+b N^{-2/4}$, yields $T_0=0.6979(2)$ for
$N\geq64$ with $Q\approx0.1$.  Both estimates differ from our previous results
outside the error bars, which is expected because the fit error underestimates
systematic uncertainties. This is partially due to the fact that we are not yet
fully in the asymptotic scaling regime, which we will discuss below. The general
trend of the results is compatible with our previous conclusion, where we
already noted that the available three-dimensional Lennard-Jones system sizes
were too small for clear results. However, it is worth noting that the current
estimates are closer to each other, implying an infinite-size limit of the
transition temperature around $T_0\approx0.70$. 

The distance from the asymptotic scaling regime is best visible in the rounding
of the transition. This is here obtained from the half width of the specific-heat
peak, i.e., the width for which $C_V(T)>C_V^{\rm max}/2$, shown in
Fig.~\ref{figFSS}~(right). Since the width vanishes in the thermodynamic limit,
a double-logarithmic plot reveals the power-law scaling Eq.~\eqref{eqFSSDT} as a
straight line, here dashed black. As previously noticed~\cite{zierenberg2015},
an intermediate scaling regime $\propto N^{-1}$ is clearly visible and marked
with a dashed-dotted gray line. This is consistent with the droplet as relevant system
size as in this regime still a large fraction of particles contribute to the
transition droplet. However, for $N\geq500$ it appears that the asymptotic
scaling behavior slowly manifests itself and leading-order estimates become more
reliable.

\newcommand{\DTansatz}{\Delta T\propto N^{-\alpha}}
This can be tested by performing a direct fit of the power-law ansatz
$\DTansatz$, shown in Fig.~\ref{figRounding} for variable lower bound $N_{\rm
min}$. Only for $N_{\rm min}\geq640$ the fit yields plausible goodness-of-fit
values $Q>0.1$. Still, the distance to the predicted asymptotic scaling regime
is noticeable. This is consistent with results obtained for the
three-dimensional lattice gas where decent results were only obtained for
$N_{\rm min}\approx2000$~\cite{zierenberg2015}. This emphasizes once more that
care has to be taken with leading-order finite-size scaling away from the
asymptotic scaling regime. While estimates of the thermodynamic limit become
better with increasing system size, fit errors are difficult to judge because
the underlying ansatz is not accurate enough. The same holds true for additional
higher-order corrections. While the results are surely good estimates, the fit
errors are a difficult measure, not capturing the uncertainty that comes from an
incomplete fit model~\cite{young2015}.

\section{Conclusions}
We have extended our previous results~\cite{zierenberg2015} for the
three-dimensional Lennard-Jones system to larger system sizes. This allows for
more consistent estimates of the thermodynamic limit when considering
leading-order and higher-order fits. Still, the current Lennard-Jones system
with up to $2048$ particles at density $\rho=10^{-2}$ is quite far away from the
asymptotic scaling limit. We notice that the rounding of the transition is a
good indicator to visualize this distance from the asymptotic scaling regime.
Here, also the emerging intermediate scaling regime is best noticeable. 

\begin{figure}
  \centering
  \includegraphics{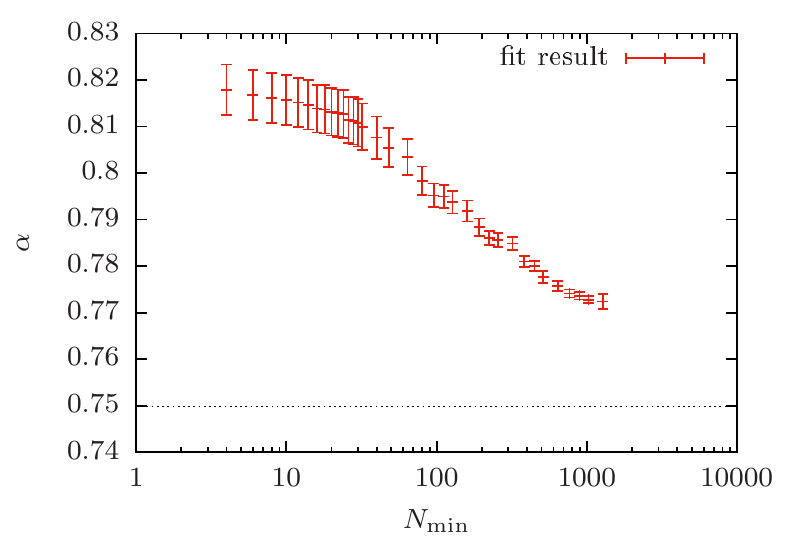}
  \caption{\label{figRounding}%
    Results of direct fits of the ansatz $\DTansatz$ to the data in
    Fig.~\ref{figFSS}~(right) with variable lower bound $N_{\rm min}$. The
    dotted line shows the predicted asymptotic scaling exponent.
  }
\end{figure}

\section*{Acknowledgments}
The project was funded by the European Union, the Free State of Saxony and the
Deutsche Forschungsgemeinschaft (DFG) under Grant No.\ JA~483/31-1. The authors
gratefully acknowledge the computing time provided by the John von Neumann
Institute for Computing (NIC) on the supercomputer JURECA at J\"ulich
Supercomputing Centre (JSC) under Grant No.\ HLZ24.  Part of this work has been
financially supported by the DFG through the Leipzig Graduate School of Natural
Sciences ``BuildMoNa'' and by the Deutsch-Franz\"osische Hochschule (DFH-UFA)
through the Doctoral College ``${\mathbb L}^4$'' under Grant No.\ CDFA-02-07.

\bibliographystyle{iopart-num}
\section*{References}
\bibliography{references}
\end{document}